\documentclass[aps,prd,preprint,superscriptaddress,floatfix,nofootinbib]{revtex4}

\usepackage[dvips]{graphicx}

\newcommand{\lsim}{\mathrel{\mathop{\kern 0pt \rlap
  {\raise.2ex\hbox{$<$}}}
  \lower.9ex\hbox{\kern-.190em $\sim$}}}
\newcommand{\gsim}{\mathrel{\mathop{\kern 0pt \rlap
  {\raise.2ex\hbox{$>$}}}
  \lower.9ex\hbox{\kern-.190em $\sim$}}}

\newcommand{\be}{\begin{equation}}
\newcommand{\ee}{\end{equation}}
\newcommand{\beqarr}{\begin{eqnarray}}
\newcommand{\eeqarr}{\end{eqnarray}}
\newcommand{\sigmav}{\langle \sigma_{\rm ann} v \rangle}

\begin{document}

\preprint{DFTT 24/2002}
\preprint{hep-ph/yymmnn}

\title{~\\ Supersymmetric Dark Matter and  \\ the Reheating  Temperature 
of the Universe}


  
%
\author{N. Fornengo} 
\email{fornengo@to.infn.it}
\homepage{http://www.to.infn.it/~fornengo}
\affiliation{Dipartimento di Fisica Teorica, Universit\`a di Torino \\
Istituto Nazionale di Fisica Nucleare, Sezione di Torino \\
via P. Giuria 1, I--10125 Torino, Italy}

\author{A. Riotto}
\email{antonio.riotto@pd.infn.it}
\affiliation{Istituto Nazionale di Fisica Nucleare, Sezione di Padova \\
via F. Marzolo 8, I--35131 Padova, Italy}

\author{S. Scopel} 
\email{scopel@to.infn.it}
\affiliation{Dipartimento di Fisica Teorica, Universit\`a di Torino \\
Istituto Nazionale di Fisica Nucleare, Sezione di Torino \\
via P. Giuria 1, I--10125 Torino, Italy}

\date{\today}

\begin{abstract} \vspace{1cm}  

Since the thermal history of the Universe is unknown before the epoch
of primordial nucleosynthesis, the largest temperature of the
radiation dominated phase (the reheating temperature) might have been
as low as 1 MeV.  We perform a quantitative study of supersymmetric
dark matter relic abundance in cosmological scenarios with low
reheating temperature. We show that, for values of the reheating
temperature smaller than about 30 GeV, the domains of the supergravity
parameter space which are compatible with the hypothesis that dark
matter is composed by neutralinos are largely enhanced. We also find a
lower bound on the reheating temperature: if the latter is smaller
than about 1 GeV neutralinos cannot be efficiently produced in the
early Universe and then they are not able to explain the present
amount of dark matter.

\end{abstract}


\maketitle

\section{Introduction}
\label{sec:intro}

Supersymmetric dark matter provides one of the hottest topics at the
border between Cosmology and Particle Physics. This is due to the fact
that in $R$--parity conserving Supersymmetric theories the Lightest
Supersymmetric Particle (LSP) is stable and may provide the cold dark
matter, whose existence is inferred by a large number of independent
observations \cite{DM,cosmo_params}.
Among the different supersymmetric candidates, the neutralino turns
out to be a perfect dark matter particle, since it has neither charge
nor colour, its only interactions being of the weak type. 

The present abundance of neutralinos depends on the thermal history of
the Universe. In the early Universe interactions may keep neutralinos
in thermal equilibrium with the radiation bath until their abundance
freezes out at a temperature $T_F$. The neutralino mass is constrained
by accelerator data to be heavier then a few tens of GeV. This implies
that it decouples in the early Universe when it is non relativistic,
at $T_F$ in the GeV range.  This picture is correct if the maximum
temperature in the radiation-dominated era, which from now on we will
refer to as the reheating temperature $T_{RH}$, is much larger than
the freeze-out temperature $T_F$. If this is the case, the neutralino
relic abundance turns out to be:
\begin{equation}
\Omega_{\chi} h^{2}\equiv \frac{\rho_{\chi}}{\rho_c}h^2 \propto
\frac{2.6 \cdot 10^{-10} {\rm GeV^{-2}}}{\langle \sigma_{\rm ann} v
\rangle}
\label{eq:omega_intro}
\end{equation}
where $\rho_c\equiv 1.8791 h^2 \times 10^{-29}$ g cm$^{-3}$ is the
critical density, $h$ is the Hubble constant in units of 100 km
sec$^{-1}$ pc$^{-1}$, while $\sigma_{\rm ann}$ is the WIMP
pair--annihilation cross section, $v$ is the relative velocity between
the two annihilating particles, and brackets denote thermal
average. Indeed, what specifically makes the neutralino an ideal dark
matter candidate is that in Eq. (\ref{eq:omega_intro}) the value of
the annihilation cross section, calculated in a wide variety of susy
models, may yield a result for $\Omega_\chi h^2$ which falls in the
correct interval suggested by present day observations for the amount
of non--baryonic dark matter in the Universe\cite{DM,cosmo_params}:
\begin{equation}
0.05 \lsim \Omega_{\rm M} h^2 \lsim 0.3.
\label{eq:omega_interval}
\end{equation} 

When exploring the neutralino parameter space and its chances of
discovery both in accelerator and non--accelerator searches, this
argument has been usually turned the other way around, and
Eq. (\ref{eq:omega_interval}) used as a constraint on the neutralino
parameter space. Depending on the particular supersymmetric scenario,
this may have important consequences on the allowed supersymmetric
configurations.  In particular, in Supergravity--inspired models
(SUGRA), the allowed neutralino parameter space turns out to be
severely reduced by requiring that $\Omega_\chi h^2$ falls inside the
interval defined by Eq. (\ref{eq:omega_interval})
\cite{Berezinsky:1995cj,Berezinsky:1996ga,noi_sugra,Ellis.HOW,Ellis.FGO,Ellis:2002wv,Ellis.more,Feng,Arnowitt,Nath,Roszkowski}.

The robustness of these constraints relies on the cosmological
assumptions that lead to Eq. ({\ref{eq:omega_intro}). Indeed, the
thermal history of the Universe before the epoch of nucleosynthesis is
unknown. The maximum temperature in the radiation-dominated era
$T_{RH}$ may have been as low as 1 MeV (but not smaller in order not
to spoil the nucleosynthesis predictions).  The possibility of a low
reheating temperature of the Universe has been recently discussed in
Ref. \cite{Giudice:2000ex}. There it was shown that a low reheating
temperature has important implications for many topics in cosmology
such as axion physics, leptogenesis and nucleosynthesis constraints on
decaying particles. In particular, it was shown that stable weakly
interacting massive particles may be produced even if the reheating
temperature is much smaller than the freeze-out temperature of the
dark matter particles, $T_{RH}< T_F$, and that the dependence of the
present abundance on the mass and the annihilation cross section of
the dark matter particle may differ drastically from standard
results\footnote{Low reheating scenarios lead as well to a new
perspective on baryogenesis \cite{Davidson:2000dw} and to the
possibility that massive neutrinos may play the role of warm dark
matter \cite{Giudice:2000dp}.}.

The goal of this paper is twofold: first, we wish to extend the
analysis of Ref. \cite{Giudice:2000ex} and perform a quantitative
study of the case of neutralinos in SUGRA scenarios, analyzing in
detail the impact that a low $T_{RH}$ may have for the present
neutralino relic abundance; secondly, we aim at providing a lower
bound on the reheating temperature. The logic is the following.  All
matter is produced at the end of inflation \cite{reviewinf} when all
the vacuum energy stored into the inflaton field is released and the
Universe becomes radiation-dominated with the initial temperature
$T_{RH}$ .  During the reheating process, particles are generated
through thermal scatterings and quickly thermalize.  Among them, dark
matter particles may be also produced but their final number depends
strongly on the reheating temperature. If the latter is too small, the
thermal bath does not give rise to a number of neutralinos large
enough to make them good candidates for dark matter. This leads to a
lower limit on $T_{RH}$. We will find that the reheating temperature
needs to be larger than about 1 GeV for neutralinos to be good dark
matter candidates\footnote{In this paper we suppose that neutralinos
are produced during the reheating process only through thermal
scatterings. Another source might be the direct decay of the inflaton
field into neutralinos \cite{Moroi:1999zb}. This introduces though
another unknown parameter, the decay rate of the inflaton field into
neutralinos, and we do not consider this possibility any further.}.

The plan of the paper is as follows. In Section II we briefly recall
the calculation of a WIMP relic abundance both in the low--reheating
scenario and in the standard case. In Section III the specific
supersymmetric models which will be considered in our analysis are
introduced. In Section IV we will discuss our results. Section V is
devoted to our conclusions.

\section{Neutralino relic abundance: the standard scenario and the
low reheating temperature scenario}
\label{sec:omega}

In this section we outline the ingredients which are relevant to the
calculation of the neutralino relic abundance both in the standard
radiation--dominated scenario and in the low reheating
early Universe scenario introduced in Ref. \cite{Giudice:2000ex}. We
address the reader to Refs. \cite{Bottino:1993zx,Giudice:2000ex} for
further details.

\subsection{The Standard Scenario}
\label{sec:omegastandard}

The number density $n_{\chi}$ of neutralinos in the early Universe is
governed by the Boltzmann equation which takes into account both the
expansion of the Universe and the neutralino interactions in the
primordial plasma:
\be 
\frac{dn_{\chi}}{dt}=-3 H n_{\chi}- \sigmav \left
[ n^2_{\chi}-(n^{\rm eq}_{\chi})^2 \right ] \, ,
\label{eq:mb_chi}
\ee 
where $H$ is the Hubble parameter, $t$ denotes time and $n^{\rm
eq}_{\chi}$ is the neutralino equilibrium number density.  In the {\it
r.h.s.} of Eq. (\ref{eq:mb_chi}), the first term describes the
Universe expansion, while the second term takes into account the
change in the $\chi$ number density due to annihilation and
inverse--annihilation processes.

At high temperatures, the evolution of $n_{\chi}$ closely tracks its
equilibrium value $n^{\rm eq}_{\chi}$. In this regime the interaction
rate of the $\chi$ particles is strong enough to keep them in thermal
equilibrium with the plasma. As temperature decreases, for heavy
particles like neutralinos the quantity $n^{\rm eq}_{\chi}$ becomes
exponentially suppressed and therefore the interaction rate $\Gamma =
n_\chi \sigmav$ turns out to be rapidly ineffective in maintaining
neutralinos in thermal equilibrium: when the $\chi$'s mean free path
becomes of the order of the Hubble scale, $\chi$'s interactions are
freezed--out and the $\chi$'s number density per comoving volume is
freezed--in. This situation occurs at a temperature $T_F$ (freeze--out
temperature) and clearly depends on the strength of $\chi$'s
interactions.

By integrating the Boltzmann equation of Eq. (\ref{eq:mb_chi}) up to
the present time, one finds the neutralino relic density
$\Omega_\chi h^2$:
\begin{equation}
\Omega_\chi h^2 = \frac{m_\chi n_\chi(T=0)}{\rho_c}\, h^2 \, .
\end{equation}

A simple analytic approximation of the solution of the Boltzmann
equation (\ref{eq:mb_chi}) allows us to write down explicitly
both the neutralino relic abundance:
\be
\Omega_{\chi} h^2 \simeq 8.77 \times 10^{-11} \,
\frac{1}{g_{*}^{1/2}(T_{F,std})} \,
\frac{\rm GeV^{-2}}{\sigmav_{\rm int}} \, .
\label{eq:omega_std}
\ee
and the value of the freeze-out temperature for the standard scenario
$T_F = T_{F,std} $ in the implicit form:
\begin{equation}
x_{F,std} \simeq \ln \left [ 0.038\, \frac{g \, m_\chi \, M_P \,
x_{F,std}^{1/2}}{g_{*}^{1/2}(T_{F,std})} \, \sigmav_F \right ]\, .
\label{eq:xf_std}
\end{equation}
where $x_{F,std} \equiv m_\chi / T_{F,std}$ and $\sigmav_F$ denotes
the value of $\sigmav$ at the freeze-out temperature. In
Eqs. (\ref{eq:omega_std},\ref{eq:xf_std}) $M_P$ denotes the Planck
mass, $g$ is the number of internal degrees of freedom of $\chi$,
$g_{*}(T_F)$ is the effective number of degrees of freedom of the
plasma at the freeze-out and $\sigmav_{\rm int}$ denotes the
integrated value of $\sigmav$ from $T_F$ up to the present
temperature. Making use of the non--relativistic first--order expansion
of $\sigmav$ in terms of the variable $x \equiv m_\chi / T $:
\begin{equation}
\sigmav = \tilde{a} +  \tilde{b} \, x^{-1} \, ,
\label{eq:expansion}
\end{equation}
we have: $\sigmav_{\rm int} = \tilde{a} \, x_F^{-1} + \tilde{b}\,
x_F^{-2}/2$. Eq. (\ref{eq:omega_std}) shows the well know result that
the present abundance of a cold relic particle is inversely
proportional to its annihilation cross section: $\Omega_\chi h^2 \sim
\sigmav_{\rm int} ^{-1}$.

A crucial point in this discussion is that, in the standard
cosmological scenario, freeze--out occurs in a phase of the evolution
of the Universe when the expansion is adiabatic and the energy
density is dominated by radiation: $T \sim a^{-1}$ and $H \sim T^2
\sim a^{-2}$. These relations between the temperature, the scale
factor $a$ and the Hubble parameter are modified in the reheating
phase of the low reheating--temperature scenario discussed in the next
Section: if freeze--out occurs during the reheating phase, a lower
neutralino relic abundance at present time is obtained.

\subsection{A low reheating--temperature scenario}
\label{sec:lowreheating}

It is by now accepted that  during the early epochs of the Universe
there was a primordial
stage of inflation \cite{reviewinf} responsible for the observed
homogeneity and isotropy of the present Universe as well as for the
generation of the cosmological perturbations.

The radiation--dominated era of the Universe is usually assumed to be
originated by the decay of the coherent oscillations of a scalar
field, the inflaton field, whose vacuum energy has driven inflation\footnote{
Identifying this scalar field with the inflaton field is not strictly 
necessary. It might well be identified with some massive nearly
stable particle, such as 
some light modulus field present in  supersymmetric and 
(super)string models. In such a case, after inflation
 the Universe might have
 been matter dominated
by the energy density of this modulus and then become
radiation dominated after its decay. In other words,
there might have been more than one reheating process during the thermal
history of the Universe. Needless to say, the one relevant for us is the
latest.}.
The decay of the scalar field into light degrees of freedom  and 
their subsequent
thermalization, called reheating, leaves the
Universe at a temperature $T_{RH}$, which represents the largest
temperature of the plasma during the subsequent radiation--dominated
epoch, when temperature is a decreasing function of time. The onset of
the radiation dominated era is in fact placed at the temperature
$T_{RH}$, {\em i.e.} at the end of the reheating phase.

Usually $T_{RH}$ is assumed to be very large and -- in any case --
larger than the neutralino freeze-out temperature $T_F$.  This fact
implies that the present--day relic abundance of any particle which
freezes--out at a temperature $T_F<T_{RH}$ is not affected by the
history of the Universe during the reheating phase. However the only
information we have on the smallest value of $T_{RH}$ is from
requiring a successful period of primordial nucleosynthesis,
$T_{RH}\gsim 1$ MeV. Therefore, from a phenomenological point of view,
$T_{RH}$ is actually a free parameter. This implies that the situation
in which a relic particle decoupled from the plasma before reheating
was completed ({\em i.e.}: $T_F>T_{RH}$) could be a viable possibility,
with important implications in the calculation of the cosmological
abundance of relic particles.

Let us consider the scenario of early Universe discussed in
Ref. \cite{Giudice:2000ex}. During the reheating epoch, the energy
density of the Universe is dominated by the coherent oscillations of a
scalar field $\phi$. This period begins at a time $H^{-1}_I$ and lasts
until a time $\Gamma_{\phi}^{-1}$ set by the scalar field decay rate
$\Gamma_{\phi}$. The dynamics of the system for
$H^{-1}_I<t<\Gamma_{\phi}^{-1}$ is described by the Boltzmann
equations for the energy densities $\rho_{\phi,R,\chi}$ of the three
coupled components: the (unstable) massive field $\phi$, radiation
$R$, and the (stable) massive WIMP $\chi$'s:
\beqarr
\frac{d \rho_{\phi}}{dt}&=&-3 H \rho_{\phi}-\Gamma_{\phi}\rho_{\phi} \, ,
\label{eq:rho_phi}\\
\frac{d \rho_{R}}{dt}&=& -4 H \rho_R + \Gamma_{\phi} \rho_{\phi} +
(2 \langle E_{\chi} \rangle) \langle \sigma_{\rm ann} v \rangle 
\left [n_{\chi}^2- (n_{\chi}^{eq})^2 \right ] \, , \label{eq:rho_R}\\
\frac{d n_{\chi}}{dt}&=& - 3 H n_{\chi} -\langle \sigma_{\rm ann} v \rangle 
\left [n_{\chi}^2-(n_{\chi}^{eq})^2 \right ] \, , \label{eq:rho_chi}
\eeqarr
where the quantity $2\langle E_{\chi} \rangle$ represents the average
energy released in each $\chi \chi$ annihilation. Notice that here we
assume that $\phi$ decays into radiation, but not into $\chi$'s.
In the following we will recall the main properties of the system
described by Eqs. (\ref{eq:rho_phi},\ref{eq:rho_R},\ref{eq:rho_chi}),
addressing the reader to Ref.\cite{Giudice:2000ex} for a complete
discussion.

At very early times the energy density of the Universe is dominated by
the scalar filed $\phi$, and both the radiation and WIMP energy
densities are negligible. As the scalar field decays the temperature
$T$ grows until it reaches a maximum value $T_{MAX}$ and then
decreases as $T\propto a^{-3/8}$ up to the temperature $T_{RH}$ at the
time $t\simeq \Gamma_{\phi}^{-1}$ which determines the end of
reheating (in this second stage the temperature of the radiation
produced in the first stage is cooled down by expansion, and the
entropy release due to the decay of $\phi$ induces a gentler cooling
compared to the radiation--dominated case). This non--standard
relation between the temperature and the scale factor may
significantly affect the calculation of the $\chi$ particles relic
abundance, depending on the duration and on the details of the
reheating phase.

The relevant mass scales which can help in understanding the different
regimes which may occur are: the mass of the WIMP $m_\chi$, the
temperature $T_F$ of the WIMP freeze--out (which can happen before or
after the reheating has been completed), the temperature at the end of
the reheating phase $T_{RH}$ and the maximal temperature reached
during the reheating phase $T_{MAX}$. Two hierarchies are present for
these mass scales: $T_{RH} < T_{MAX}$ and, since we are dealing with
cold relics which decouple when non--relativistic, $T_F < m_\chi$.

The relation between $m_\chi$ and $T_{MAX}$ determines whether the
WIMPs, which are produced during the reheating phase, become
relativistic ($T_{MAX} > m_\chi$) or not ($T_{MAX} < m_\chi$). More
important is to determine whether the $\chi$ particles reach thermal
equilibrium during the reheating phase. This is determined by the
strength of their interactions, and in particular it depends on
$\sigmav$. According to the values of the WIMP mass and annihilation
cross section, two possible non--standard regimes may be schematically
singled out for the WIMP relic abundance:

{\em (i)} The $\chi$ particles generated during reheating are always
non--relativistic and never reach thermal equilibrium. Integration of
Eqs.(\ref{eq:rho_phi},\ref{eq:rho_R},\ref{eq:rho_chi}) shows that the
process of $\chi$ production takes its main contribution around the
temperature $T_{*}\simeq m_{\chi}/4$ and is exponentially suppressed
outside a narrow interval centered on $T_{*}$
\cite{Giudice:2000ex}. So most of the $\chi$ particles are produced at
$T_{*}$. For $T<T_{*}$ the total number of $\chi$'s is frozen and
their density is diluted by expansion. The condition $n_{\chi}(T=0)
<n_{\chi}^{eq} (T_*)$ implies an upper limit on $\sigmav_* \equiv
\sigmav_{(T=T_*)}$ of the order of \cite{Giudice:2000ex}:
\be \langle \sigma_{ann} v \rangle_* \lsim
7\times 10^{-14} \, \frac{2}{g} \left [\frac{g_*(T_*)}{10} \right ]
\left [\frac{10}{g_*(T_{RH})} \right ]^{1/2}
\left ( \frac{m_{\chi}}{100 \; \rm GeV} \right )
\left ( \frac{100 \; \rm MeV}{T_{RH}} \right )^2 \; {\rm GeV^{-2}}.
\label{eq:sigma_star}
\ee
In this regime the WIMP relic density turns out to be {\em
proportional} to the WIMP self--annihilation cross section:
$\Omega_\chi h^2 \sim \sigmav_*$, instead of being inversely
proportional, as in the standard case. We anticipate here that for the
case of neutralinos in supersymmetric models, this situation only
occurs when the reheating temperature is smaller than about 300
MeV. In this situation the neutralino relic density is significantly
suppressed.

{\em (ii)} The $\chi$ particles reach thermal equilibrium, and then
freeze--out when non--relativistic before the reheating phase is
concluded, {\em i.e.} at a temperature $T_F = T_{F,rh} > T_{\rm
RH}$. During the phase when they reach thermal equilibrium, the $\chi$
particles may or may not become relativistic, depending on the value
of $T_{MAX}$. In any case, they decouple as non--relativistic, leading
to a cold relic. The usual freeze--out condition is modified compared
to the standard case because the energy density is dominated by the
scalar field and the relation between the Hubble constant and the
temperature is given by $H \propto T^4$, as compared to the
radiation--dominated case where $H \propto T^2$. The first consequence
of this fact is that freeze--out occurs earlier and the WIMP density
at $T_{F,rh}$ turns out to be higher. However, as the Universe cools
down from $T_{F,rh}$ to $T_{RH}$, due to the entropy produced by the
$\phi$ decays, the expansion dilutes $n_{\chi}$ by a factor
$(T_{RH}/T_{F,rh})^8$, which is much smaller than the dilution factor
in the radiation--dominate case $(T_{RH}/T_{F,std})^3$.  When the two
effects are combined, the final result is a suppression of
$\Omega_{\chi}$ by roughly a factor $T^3_{RH} T_{F,std}/(T_{F,rh})^4$
as compared to the standard case. An analytic approximation of the
ensuing result for the $\chi$ relic abundance is given by
\cite{Giudice:2000ex}:
\be \Omega_\chi h^2 \simeq 2.3 \times 10^{-11}
\frac{g_{*}^{1/2}(T_{RH})} {g_{*}(T_{F,rh})} \frac{T^3_{RH} {\rm
GeV}^{-2}}{m_{\chi}^3 (\tilde{a} x_{F,rh}^{-4}+4 \tilde{b}
x_{F,rh}^{-5}/5)} \, ,
\label{eq:omegalowreheating2}
\ee
where the freeze--out temperature during the reheating--phase
$T_{F,rh}$ ($x_{F,rh} \equiv m_\chi/T_{F,rh}$) is
\cite{Giudice:2000ex}:
\begin{equation}
x_{F,rh} \simeq \ln \left [ 0.015 \, \frac{g\,
g_*^{1/2}(T_{RH})}{g_*(T_{F,rh})} \frac{M_P \,
T_{RH}^2}{m_\chi}\, (\tilde a\, x_{F,rh}^{5/2} + 5 \tilde b
x_{F,rh}^{3/2}/4) \right ]\, .
\label{eq:xf_RH}
\end{equation}
In both Eqs. (\ref{eq:omegalowreheating2}) and (\ref{eq:xf_RH}) we have
used the annihilation cross--section expansion of Eq. (\ref{eq:expansion}).

Obviously, if the decoupling of the $\chi$ particles from the plasma
occurs after the reheating phase is concluded, the standard scenario
is recovered and the relic abundance has the ordinary expression of
Eq. (\ref{eq:omega_std}).

In the following Sections we will perform a detailed calculation of
the neutralino relic abundance in a Supergravity framework in order to
study the consequences of the low reheating temperature scenario
outlined above. In our analysis we will make use of the analytical
solutions given in Eqs.(\ref{eq:omega_std},\ref{eq:xf_std}) and
Eqs.(\ref{eq:omegalowreheating2},\ref{eq:xf_RH}), appropriately
interpolated in the intermediate regime where $T_F$ is close to
$T_{RH}$ (the interpolation has been determined on the basis of
numerical solutions of the relevant differential equations). We have
numerically verified that the analytical solutions are accurate enough
for our purposes when compared with full numerical solution of the
relevant Boltzmann equations of Eq. (\ref{eq:mb_chi}) and
Eqs. (\ref{eq:rho_phi},\ref{eq:rho_R},\ref{eq:rho_chi}).

\section{The neutralino in Minimal Supergravity}
\label{sec:SUGRA}
 
Supersymmetric theories naturally predict the existence of viable dark
matter candidates if $R$--parity is conserved, since this symmetry
prevents the lightest supersymmetric particle (LSP) from decaying. The
nature and the properties of the LSP depend on the way supersymmetry
is broken. In models where supersymmetry breaking is realized through
gravity-- (or also anomaly--) mediated mechanisms, the LSP turns out
to be quite naturally the neutralino, defined as the lowest--mass
linear superposition of photino ($\tilde \gamma$), zino ($\tilde Z$)
and the two higgsino states ($\tilde H_1^{\circ}$, $\tilde
H_2^{\circ}$): $\chi \equiv a_1 \tilde \gamma + a_2 \tilde Z + a_3
\tilde H_1^{\circ} + a_4 \tilde H_2^{\circ}$.

Even assuming a minimal supersymmetric extension of the Standard
Model, the supersymmetric theories may be explored in a variety of
different schemes, ranging from those based on universal
\cite{Berezinsky:1995cj,Ellis.HOW,Berezinsky:1996ga,noi_sugra,Ellis.FGO,Ellis.more,Feng,Arnowitt,Nath,Roszkowski}
or non--universal
\cite{Berezinsky:1995cj,Berezinsky:1996ga,noi_sugra,Arnowitt,Nath,Ellis:2002wv}
Supergravity, where the relevant independent supersymmetric parameters
are defined at a grand unification scale, to effective supersymmetric
theories which are defined at the electroweak scale
\cite{noi_sugra,noi_mssm,effMSSM}. In the present paper we will mostly
concentrate on the discussion of universal Supergravity and we will
comment on the results which can be obtained in different
supersymmetric schemes.

The essential elements of a generic minimal supersymmetric model are
described by a Yang--Mills Lagrangian, by the superpotential, which
contains all the Yukawa interactions between the standard and
supersymmetric fields, and by the soft--breaking Lagrangian, which
models the breaking of supersymmetry. Implementation of this model
within a Supergravity scheme leads quite naturally to a set of
unification conditions at a grand unification scale ($M_{GUT}$) for
the parameters of the theory:

\begin{itemize}

\item Unification of the gaugino masses: 
\begin{equation}
M_i(M_{GUT}) \equiv m_{1/2} \, , 
\label{eq:M12}
\end{equation}

\item Universality of the scalar masses with a common mass denoted by $m_0$: 
\begin{equation}
m_i(M_{GUT}) \equiv m_0 \, ,
\label{eq:m0univ}
\end{equation}
\item Universality of the trilinear scalar couplings:
\begin{equation}
A^{l}(M_{GUT}) = A^{d}(M_{GUT}) = A^{u}(M_{GUT}) \equiv A_0 m_0 \, .
\label{eq:Atrilinear}
\end{equation}

\end{itemize}
    
We denote this scheme as universal SUGRA (or minimal SUGRA). The
relevant parameters of the model at the electroweak scale are obtained
from their corresponding values at the $M_{GUT}$ scale by running
these down according to renormalization group equations. By requiring
that the electroweak symmetry breaking is induced radiatively by the
soft supersymmetry breaking, one finally reduces the parameters of the
model to five: $m_{1/2}$, $m_0$, $A_0$, $\tan \beta \equiv v_2/v_1$
and ${\rm sign}(\mu)$, where $v_1$ and $v_2$ denote the vev's of the two
Higgs field of the model and $\mu$ is a mixing parameter between the
two Higgs fields which enters in the superpotential. These parameters
are {\em a priori} undetermined. However, bounds coming from
supersymmetry and Higgs searches at accelerators and on
supersymmetric contributions to rare process, like the $b \rightarrow
s + \gamma$ radiative decay, introduce limits on the model
parameter space. Also theoretical arguments concerning the naturalness
of the theory may be used in order to identify typical scales beyond
which the main attractive features of supersymmetry fade away. For
instance, fine-tuning arguments may be invoked to set bounds on $m_0$
and $m_{1/2}$ \cite{Berezinsky:1995cj,Berezinsky:1996ga}: $m_{1/2}
\lsim$ hundreds of GeV, whereas $m_0 \lsim (2-3)$ TeV.

In the present paper we will vary the parameters of the minimal SUGRA
scheme in wide ranges in order to carefully analyze the impact of the
low reheating--temperature cosmology on relic neutralinos. The ranges
we adopt are the following: 
\begin{eqnarray}
50\;\mbox{GeV} \leq &m_{1/2}& \leq 3\;\mbox{TeV} \, , \nonumber \\
&m_0& \leq 3\;\mbox{TeV} \, , \\
-3 \leq &A_0& \leq +3 \, , \nonumber \\
1 \leq &\tan\beta& \leq 60 \, . \nonumber 
\label{eq:ranges}
\end{eqnarray}
The sign of $\mu$ is chosen to be positive, since negative values are
somehow disfavoured by the limits on the supersymmetric contribution
to the muon anomalous magnetic moment.  On the configurations obtained
by randomly scanning the above defined parameter space, we apply the
experimental limits quoted above on Higgs and supersymmetry searches
and on the $b \rightarrow s + \gamma$ decay (for details, see for
instance Ref.\cite{details,taup01}).

We finally remark that the phenomenology of relic neutralinos, in some
sectors of the minimal SUGRA scheme, is also quite sensitive to some
Standard Model parameters, such as the top quark mass $m_t$, the
bottom quark mass $m_b$ and the strong coupling $\alpha_s$
\cite{taup01,marinadelrey01,Ellis.HOW}. For these parameters, we use
here their 95\% CL ranges: $m_t^{\rm pole} = (175 \pm 10)$ GeV,
$m_b(M_Z) = (3.02 \pm 0.21)$ GeV and $\alpha_s(M_Z) = 0.118 \pm
0.004$.

\section{Results and Discussion}
\label{sec:results}

As already discussed in Sec.\ref{sec:omega}, the calculation of the
neutralino relic abundance $\Omega_{\chi} h^2$, both in the standard
case outlined in Sec.\ref{sec:omegastandard} and in the low--reheating
scenario described in Sec.\ref{sec:lowreheating}, relies on a detailed
calculation of the neutralino self--annihilation cross section
$\sigmav$. Here we consider the universal SUGRA model outlined in the
previous Section, and calculate $\sigmav$ following the procedure
given in Ref. \cite{Bottino:1993zx}, to which we refer for details.

A first preliminary conclusion about neutralino dark matter and models
with a low--reheating temperature may be drawn by calculating the
neutralino annihilation cross section at the temperature $T_*$. The
quantity $\sigmav_*$ is shown in Fig.\ref{fig:sigmavstar} as a
function of $m_{\chi}$ in terms of a scatter plot obtained by varying
the SUGRA parameters in the ranges given by Eq. (\ref{eq:ranges}). The
scatter plot is compared to the values of the limiting cross section
given in Eq. (\ref{eq:sigma_star}), calculated for different values of
$T_{RH}$. This figure shows that the condition of non--relativistic
non--equilibrium, given by Eq. (\ref{eq:sigma_star}), is verified only
for values of $T_{RH}$ smaller than about 300 MeV. For larger values
of the reheating temperature, $\sigmav_*$ always lies above the curves
of the limiting cross section. This fact implies that for $T_{RH}
\gsim 300$ MeV the neutralino always reaches thermal equilibrium
during the reheating phase, and therefore the peculiar behaviour
$\Omega_\chi h^2 \propto \sigmav$ is limited only to cosmological
models with very low reheating temperatures. This conclusion is
actually true for a class of supersymetric scenarios which is more
general than the universal SUGRA model shown in
Fig.\ref{fig:sigmavstar}. For instance, we have explicitly verified
that the same result also applies to other SUGRA schemes where the
universality condition of Eq. (\ref{eq:m0univ}) are relaxed at the GUT
scale for the Higgs sector, when the supersymmetric parameters are
varied in the same ranges discussed in the previous Section. One must
notice that the effective lower limit on neutralino cross section
revealed in Fig. \ref{fig:sigmavstar} is a consequence of the choice
of the upper ranges of Eq. (\ref{eq:ranges}) adopted for the
dimensional supersymmetric parameters. However, we remind that the
intervals given in Eq. (\ref{eq:ranges}) are representative of a
typical upper bound on the supersymmetry breaking scale, $M_{\rm susy}
\lsim {\rm a\; few}\times 10^3$ GeV. This constraint may be understood
on quite general grounds, since it derives from naturalness arguments
on the stability of the Higgs potential and the requirement of absence
of fine--tuning in the generation of the electroweak scale through the
mechanism of radiative symmetry breaking. Notice that, as far as
$m_{1/2}$ is concerned, the upper value adopted in
Eq. (\ref{eq:ranges}) is already large enough to pose questions about
fine--tuning problems \cite{Berezinsky:1995cj,Berezinsky:1996ga}.

Let us turn now to the calculation of the neutralino relic abundance.
Fig.\ref{fig:omega} shows $\Omega_\chi h^2$ as a function of the
neutralino mass $m_\chi$, calculated in universal SUGRA for $\tan\beta
= 30$ and $A_0=0$ and with $m_0$ and $m_{1/2}$ varied in the ranges of
Eq. (\ref{eq:ranges}). Starting from the upper--left panel, we plot
the results obtained for standard cosmology, and for low
reheating--temperature cosmologies with $T_{RH}=30$, 20, 10, 5, 1
GeV. In the case of standard cosmology, the neutralino relic abundance
turns out to be generically quite large, in excess of the upper bound
on the total amount of matter in the Universe given in
Eq. (\ref{eq:omega_interval}). This is a typical feature of universal
SUGRA models for values of $\tan\beta \lsim 40$
\cite{Berezinsky:1995cj,Berezinsky:1996ga,noi_sugra,Ellis.HOW,Ellis.FGO,Ellis.more,Feng,Arnowitt,Nath,Roszkowski}. For
this schemes, the constraint on the parameter space coming from
cosmology is actually very strong, especially in posing stringent
upper limits on the neutralino mass. Fig. \ref{fig:omega} shows that
an upper bound of about 200 GeV is obtained on the neutralino mass
when we restrict $\Omega_\chi h^2$ to be less than 0.3.

There are, however, ways out to avoid such a bound. When
coannihilation is included, the upper bound on $m_\chi$ in universal
SUGRA at low/intermediate $\tan\beta$ can be extended up to about 500
GeV \cite{Ellis.HOW,Ellis.FGO,Arnowitt,Roszkowski}, even though this
possibility is restricted to a very narrow sector of the SUGRA
parameter space where the neutralino mass is almost degenerate with
the stau mass.

Alternatively, when a low reheating temperature is allowed,
Fig.\ref{fig:omega} shows that the upper limit on $m_\chi$ coming from
cosmology is removed. By lowering the reheating temperature we affect
mainly the relic abundance for large neutralino masses, unless we
lower $T_{RH}$ below a few GeV: for $T_{RH} \lsim 5$ GeV, $\Omega_\chi
h^2$ is reduced for all the allowed mass range of the SUGRA model.

The behaviour of the different panels of Fig.\ref{fig:omega} may be
easily understood by comparing the relevant mass scales which enter in
the calculation of the relic abundance in the standard and low
reheating--temperature models, namely $m_\chi$, $T_F$ and $T_{RH}$.
In Fig.\ref{fig:omega_one} we show a generic example of what happens
at different neutralino masses for a sufficiently low reheating
temperature. The upper panel of Fig.\ref{fig:omega_one} shows the
relic abundance vs. $m_\chi$, calculated for $\sigmav$ fixed at the
value $10^{-10}$ GeV$^{-2}$. The thick horizontal (black) line refers
to the calculation for standard cosmology, while the thick decreasing
(blue) line refers to $T_{RH} = 10$ GeV. The two lines are
superimposed for $m_\chi \lsim 150$ GeV. The lower panel shows the
value of the freeze--out temperature $T_F$ as a function of the
neutralino mass. The horizontal line indicates the values of $T_{RH} =
10$ GeV. By comparing the upper an lower panels, we see that as long
as the freeze--out occurs at a temperature which is (much) smaller
than $T_{RH}$, no difference is present for the low
reheating--temperature scenario as compared to the standard case.  On
the contrary, when $T_F$ approaches $T_{RH}$, the relic abundance
becomes suppressed, and this effect becomes more pronounced as $T_F$
grows as compared to $T_{RH}$. Since $T_F$ is an increasing function
of the neutralino mass, a low--reheating scenario affects more the
large mass sector of the theory, and this explains the peculiar
behaviour of Fig. \ref{fig:omega}.

Going back to the discussion of the features of Fig. \ref{fig:omega},
we see that in order to have heavy neutralinos compatible with the
cosmological upper bound in Eq. (\ref{eq:omega_interval}), the
reheating temperature has to be lowered below about 30 GeV. When
$T_{RH}$ falls below 20 GeV all the neutralino mass range is
acceptable from the point of view of cosmology, although a fraction of
the SUGRA configurations entail relic neutralinos which are a
subdominant dark matter component, since $\Omega_\chi h^2$ falls below
the lower limit of Eq. (\ref{eq:omega_interval}). When $T_{RH}
\lsim 0.6$ GeV all the SUGRA configurations lead to subdominant relic
neutralinos. For smaller values of $T_{RH}$ the neutralino relic
abundance is even more suppressed and becomes negligible for $T_{RH}$
in the MeV range.

We comment at this point that for larger values of $\tan\beta$
($\tan\beta \gsim 40$) in SUGRA models an upper limit on the
neutralino mass is not present even in standard cosmology. The
occurrence of an acceptable relic abundance is generically confined to
corridors in the parameter space where either the relic abundance is
suppressed by coannihilation or the neutralino mass lies close to the
pole of the annihilation cross section mediated by the pseudoscalar
higgs field $A$. In a low reheating--temperature cosmology, these
features are relaxed and conclusions similar to those discussed above
in connection with Fig.\ref{fig:omega} are present. In particular, a
scale of about 1 GeV as a lower limit on $T_{RH}$ in order to have
dominant relic neutralinos is present for all the values of
$\tan\beta$ given in Eq. (\ref{eq:ranges}).

New features in the parameter space are also present when the
universality is not extended to Higgs masses
\cite{Berezinsky:1995cj,Berezinsky:1996ga,Arnowitt,Ellis:2002wv}. Effects
of non--universality modify the low--energy sector of the theory,
through the RGE evolution and the conditions of radiative electroweak
symmetry breaking: this alters the neutralino couplings and the mass
spectrum of sparticles and induces variations of the neutralino
annihilation cross section. The consequence is a change and an
extension of the regions of parameter space which are compatible with
a relic neutralino, also in standard cosmology. A low reheating
temperature has again the effect of enlarging these cosmologically
relevant sectors of the parameter space. A lower limit of about 1 GeV
on $T_{RH}$ in order to have dominant relic neutralinos is again
recovered also in non--universal SUGRA models.

Fig. \ref{fig:m0_m12} shows the same information contained in
Fig. \ref{fig:omega}, expressed in the plane $m_0$--$m_{1/2}$. The
upper--left panel refers to standard cosmology, the other panels refer
to low reheating--temperature cosmologies with $T_{RH}=30$, 20, 10, 5,
1 GeV. The value of $\tan\beta$ is fixed at 30 and $A_0=0$.  The
light-shaded (yellow) regions are excluded domains: the ones on the
left side of the $m_0$--$m_{1/2}$ plane are excluded by the
experimental bound discussed in the previous Section or by the
non--occurrence of radiative electroweak symmetry breaking, while the
ones on the lower part of the plane do not correspond to viable LSP
neutralino models. The dark--shaded (blue) areas correspond to the
domains where the neutralino relic abundance falls inside the
cosmologically relevant range of Eq. (\ref{eq:omega_interval}). The
hatched (red) regions correspond to neutralinos with a subdominant
relic abundance, {\em i.e.} to $\Omega_\chi h^2 < 0.05$.

The first panel shows that for standard cosmology the region of the
$m_0$--$m_{1/2}$ parameter space which is allowed by cosmology is
quite restricted, as discussed above. The upper limit on the
acceptable values of the neutralino relic abundance poses severe
bounds on both $m_0$ and $m_{1/2}$, of the order of a few hundreds of
GeV. When coannihilation is included
\cite{Ellis.HOW,Ellis.FGO,Arnowitt,Nath,Roszkowski}, the allowed
regions is extended in a thin band close to the lower excluded
area. However, as is evident from the other panels, when the reheating
temperature is lowered the consequent suppression on $\Omega_\chi h^2$
weakens considerably the constraints on $m_0$ and $m_{1/2}$. In
particular, the regions of the SUGRA parameter space which are
compatible with the assumption of a dominant neutralino dark matter
change significantly, depending on the actual value of $T_{RH}$. For
values of $T_{RH}$ below about 20 GeV a large fraction of the
parameter space is allowed by cosmology and for $T_{RH} \lsim 1$ GeV
all the supersymmetric parameter space is compatible with the
cosmological abundance of relic neutralinos, even though for most of
the values of the parameters the relic abundance falls below the
interval of Eq. (\ref{eq:omega_interval}).

As we have discussed before, the extension of the cosmologically
allowed regions in the plane $m_0$--$m_{1/2}$ is enlarged, in standard
cosmology, when $\tan\beta$ is larger than about 40, since in this
case coannihilation or annihilation through the $A$--pole is effective
in reducing the values of $\Omega_\chi h^2$. In the latter case, an
almost diagonal allowed band opens up
\cite{marinadelrey01,Ellis.HOW}. Larger regions are also allowed in
non--universal SUGRA models. In both cases, a low reheating
temperature has again the effect of widely enlarging the
cosmologically relevant domains in the plane $m_0$--$m_{1/2}$. We
obtain, also for the SUGRA models with $\tan\beta \gsim 40$ and for
the non--universal SUGRA schemes, that in all the supersymmetric
parameter space the neutralino relic abundance is compatible with the
range of Eq. (\ref{eq:omega_interval}) when $T_{RH} \lsim 1$ GeV.

In the above discussion, we noticed that, for any given neutralino
mass, low enough values of $T_{RH}$ would imply a suppression of
$\Omega_{\chi} h^2$ too strong to be compatible with the hypothesis of
dominant neutralino dark matter. We can therefore use this argument to
obtain a lower limit on $T_{RH}$ {\em under the assumption} that the
neutralino represents the dominant component of dark matter in the
Universe. This lower bound on $T_{RH}$ is shown as a solid line in
Fig.\ref{fig:trh_lim}, as a function of $m_{\chi}$.  To derive this
limit we have varied all the SUGRA parameters in the intervals of
Eq. (\ref{eq:ranges}). The result plotted in Fig. \ref{fig:trh_lim}
shows that if we require the neutralino to be the dominant dark matter
component, the reheating temperature in the early Universe cannot be
lower than a value ranging from 0.6 GeV up to about 20 GeV, depending
on the value of the neutralino mass. This bound on $T_{RH}$ at the GeV
scale is quite interesting, since it is sizably stronger than the
constraint given by nucleosynthesis.  We have verified that the same
result remains valid also for non--universal SUGRA models. Therefore a
lower limit of about 1 GeV on $T_{RH}$, in order to explain the dark
matter content of the Universe in terms of relic neutralinos, is a
specific feature of supergravity models.

\section{Conclusions}
\label{sec:conclusions}

In standard cosmology it is usually assumed that the temperature
$T_{RH}$ of the Universe at the beginning of the radiation--dominated
era is much higher than the supersymmetry breaking
 scale. Moreover, neutralinos decouple
from the thermal bath after the reheating phase, which followed the
end of inflation, has terminated. Under these assumptions, the allowed
parameter space of SUGRA models turns out to be severely constrained
by the requirement that the neutralino relic density does not exceed
the maximal value of the matter density of the Universe deduced from
observations.

In this paper we have performed a quantitative study of the neutralino
relic abundance in cosmological scenarios with a low reheating temperature
\cite{Giudice:2000ex}. This is a
viable possibility since the only robust lower bound on the reheating
temperature $T_{RH}$ can be set at the MeV scale, in order not to
spoil nucleosynthesis predictions. The suppression on $\Omega_\chi
h^2$ is originated by the fact that neutralinos decouple from the
thermal bath before the end of the reheating phase. In this case the
neutralino number density is diluted by entropy production and by a
higher expansion rate than in the radiation--dominated era.

For values of $T_{RH} \lsim 30$ GeV the domains of the SUGRA
parameter space which are compatible with dominant relic neutralinos
are largely enhanced with respect to the standard cosmological
case. These domains depend on $T_{RH}$ and we have shown their
evolution as a function of the reheating temperature. For $T_{RH}
\lsim 1$ GeV all the SUGRA parameter space becomes compatible with the
bounds on the dark matter relic abundance (even though the neutralino
relic abundance for these low values of $T_{RH}$ is strongly
suppressed).

Since lower $T_{RH}$ imply smaller relic densities, the assumption
that neutralinos provide a major contribution to the dark matter of
the Universe implies a lower limit on $T_{RH}$. This constraint ranges
from 0.6 GeV for neutralino masses of the order of few tens of GeV, up
to 20 GeV for neutralino masses in the TeV range. This bound on
$T_{RH}$, subject to the request of explaining the dark matter content
of the Universe only in terms of relic neutralinos in SUGRA schemes,
is much stronger than the limit on $T_{RH}$ coming from
nucleosynthesis. Similar conclusions occur also for non--universal
SUGRA models.

\acknowledgements
We wish to thank G.F. Giudice for useful discussions.

\clearpage
\begin{figure} \centering
\vspace{-50pt}
\includegraphics[width=1.0\textwidth]{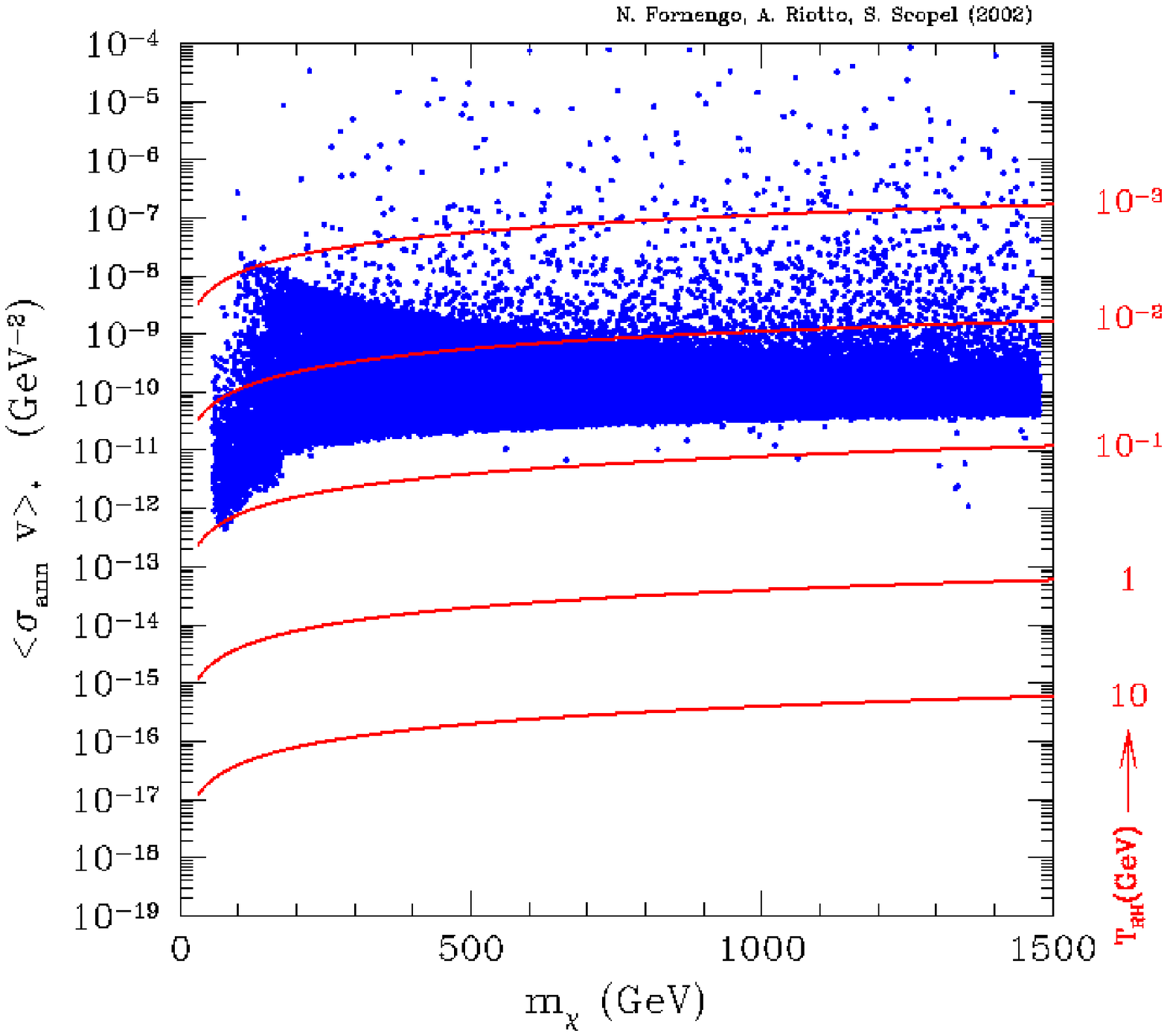}
\vspace{-50pt}
  \caption{Thermal average, at the temperature $T_{*}\sim m_{\chi}/4$,
  of the neutralino self--annihilation cross section times the
  relative velocity $\sigmav_\star$ as a function of the neutralino
  mass $m_\chi$. The points denote the values of $\sigmav_\star$
  calculated in universal SUGRA with the parameters varied as in
  Eq. (\ref{eq:ranges}). The values of $m_t^{\rm pole}$, $m_b$ and
  $\alpha_s$ are varied inside their $2\sigma$ allowed intervals. The
  solid lines denote, for different values of the reheating
  temperature $T_{RH}$, the values of the limiting cross section of
  Eq. (\ref{eq:sigma_star}) which determines, for a cosmological model
  where neutralinos are always non--relativistic, whether neutralinos
  can reach thermal equilibrium during the reheating phase.
\label{fig:sigmavstar}} 
\end{figure}

\clearpage
\begin{figure} \centering
\vspace{-50pt}
\includegraphics[width=1.0\textwidth]{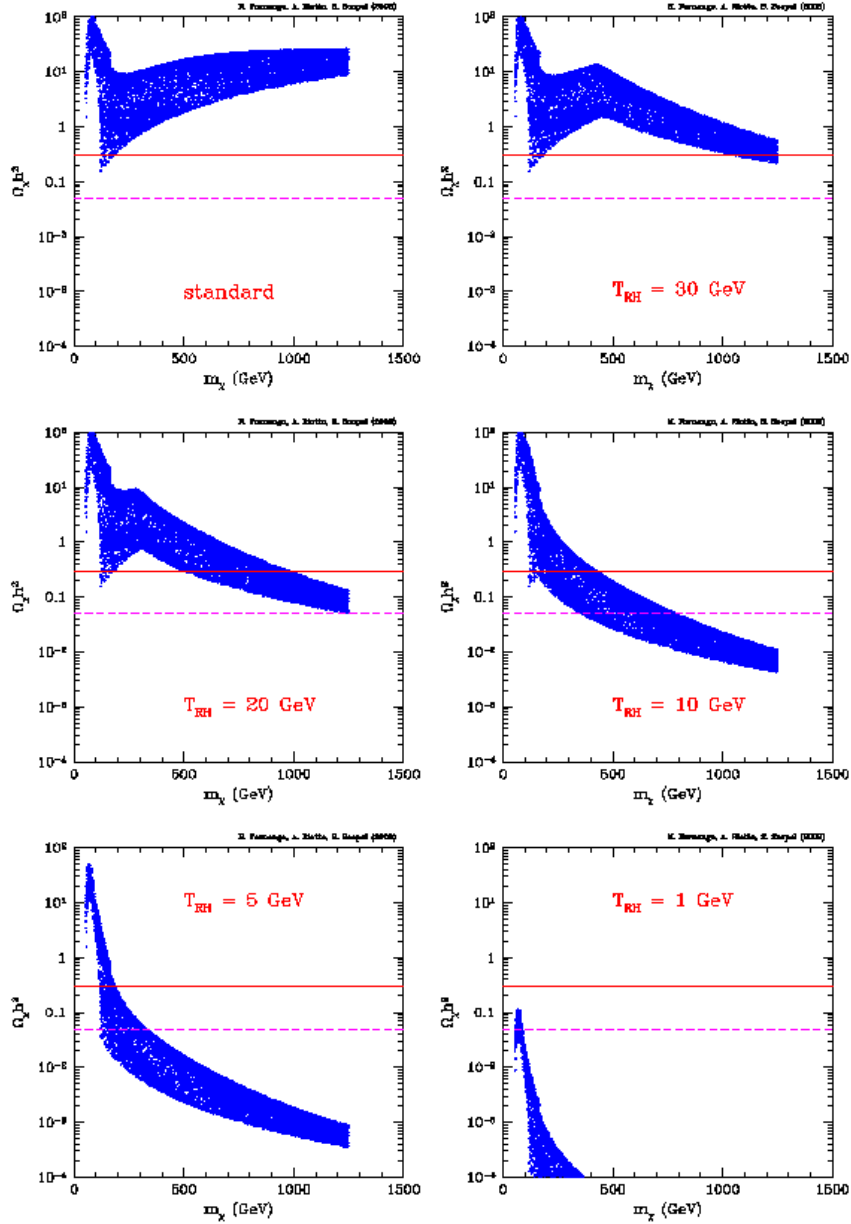}
\vspace{-100pt}
  \caption{Neutralino relic density $\Omega_\chi h^2$ as a function of
  the neutralino mass $m_\chi$ in universal SUGRA, for $\tan\beta=30$
  and $A_0=0$. The parameters $m_0$ and $m_{1/2}$ are varied according
  to the intervals of Eq. (\ref{eq:ranges}). The values of $m_t^{\rm
  pole}$, $m_b$ and $\alpha_s$ are fixed at their central values:
  $m_t^{\rm pole} = 175$ GeV, $m_b(M_Z) = 3.02$ GeV and $\alpha_s(M_Z)
  = 0.118$. The upper--left panel shows results for standard
  cosmology. The other panels refer to different values of the
  reheating temperature $T_{\rm RH}$. The horizontal solid and dashed
  lines delimit the interval for the amount of non--baryonic dark
  matter in the Universe, given in Eq. (\ref{eq:omega_interval}).
\label{fig:omega}}
\end{figure}

\clearpage
\begin{figure} \centering
\vspace{-50pt}
\includegraphics[width=0.9\textwidth]{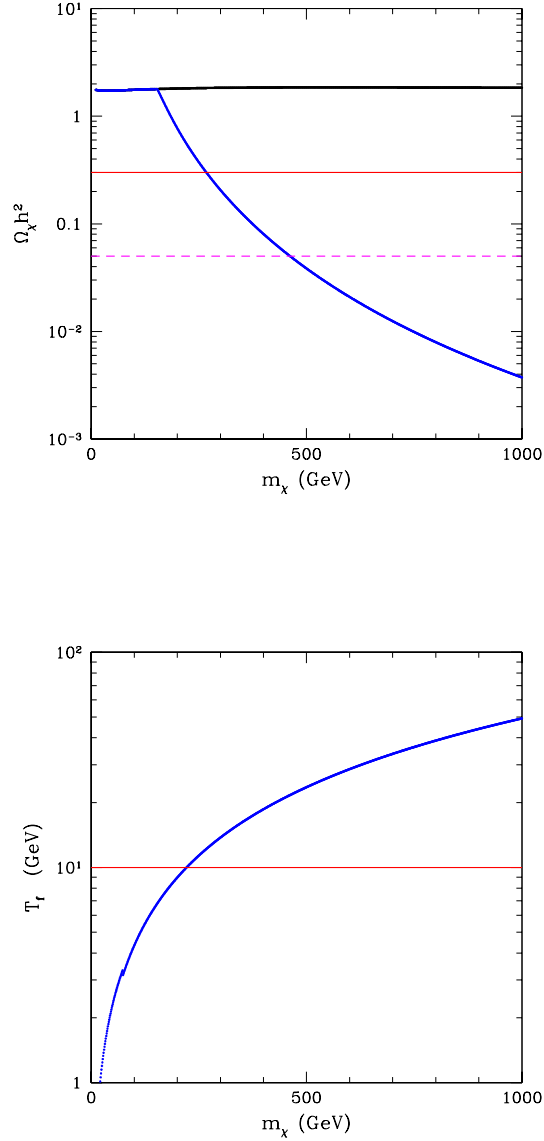} 
\vspace{-50pt}
\caption{Effect of a
  low reheating--temperature on the WIMP relic abundance. The upper
  panel shows $\Omega_\chi h^2$ as a function of $m_\chi$ for standard
  cosmology (upper thick horizontal line) and for $T_{RH}=10$ GeV
  (thick decreasing line). The two lines are superimposed for $m_\chi
  \lsim 150$ GeV.  The thin horizontal solid and dashed lines delimit the
  interval for the amount of non--baryonic dark matter in the
  Universe, given in Eq. (\ref{eq:omega_interval}). The lower panel
  shows the freeze--out temperature $T_F$ as a function of
  $m_\chi$. The horizontal line denotes the values of the reheating
  temperature: $T_{RH}=10$ GeV.
\label{fig:omega_one}}
\end{figure}

\clearpage
\begin{figure} \centering
\vspace{-50pt}
\includegraphics[width=1.0\textwidth]{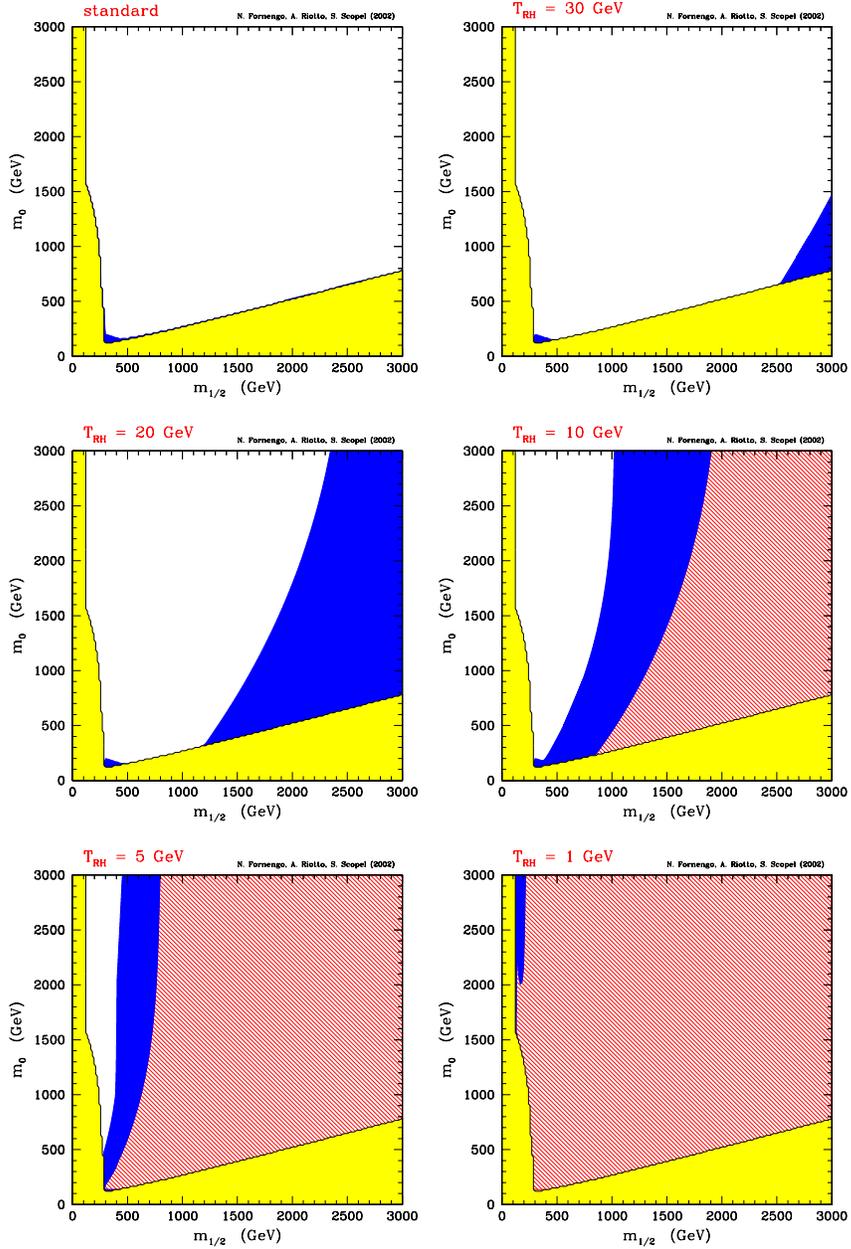}
\vspace{-100pt}
  \caption{Cosmologically favoured regions in the plane
  $m_{1/2}$--$m_0$ in universal SUGRA for $\tan\beta=30$ and
  $A_0=0$. The values of $m_t^{\rm pole}$, $m_b$ and $\alpha_s$ are
  fixed at their central values: $m_t^{\rm pole} = 175$ GeV, $m_b(M_Z)
  = 3.02$ GeV and $\alpha_s(M_Z) = 0.118$. The upper--left panel shows
  results for standard cosmology. The other panels refer to different
  values of the reheating temperature $T_{RH}$. The dark--shaded
  (blue) areas correspond to configurations where $0.05 \leq
  \Omega_\chi h^2 \leq 0.3$. Hatched (red) areas denote configurations
  where $\Omega_\chi h^2 < 0.05$. Light--shaded (yellow) regions are
  excluded either by experimental constraints or by theoretical
  arguments.
\label{fig:m0_m12}}
\end{figure}

\clearpage
\begin{figure} \centering
\vspace{-50pt}
\includegraphics[width=1.0\textwidth]{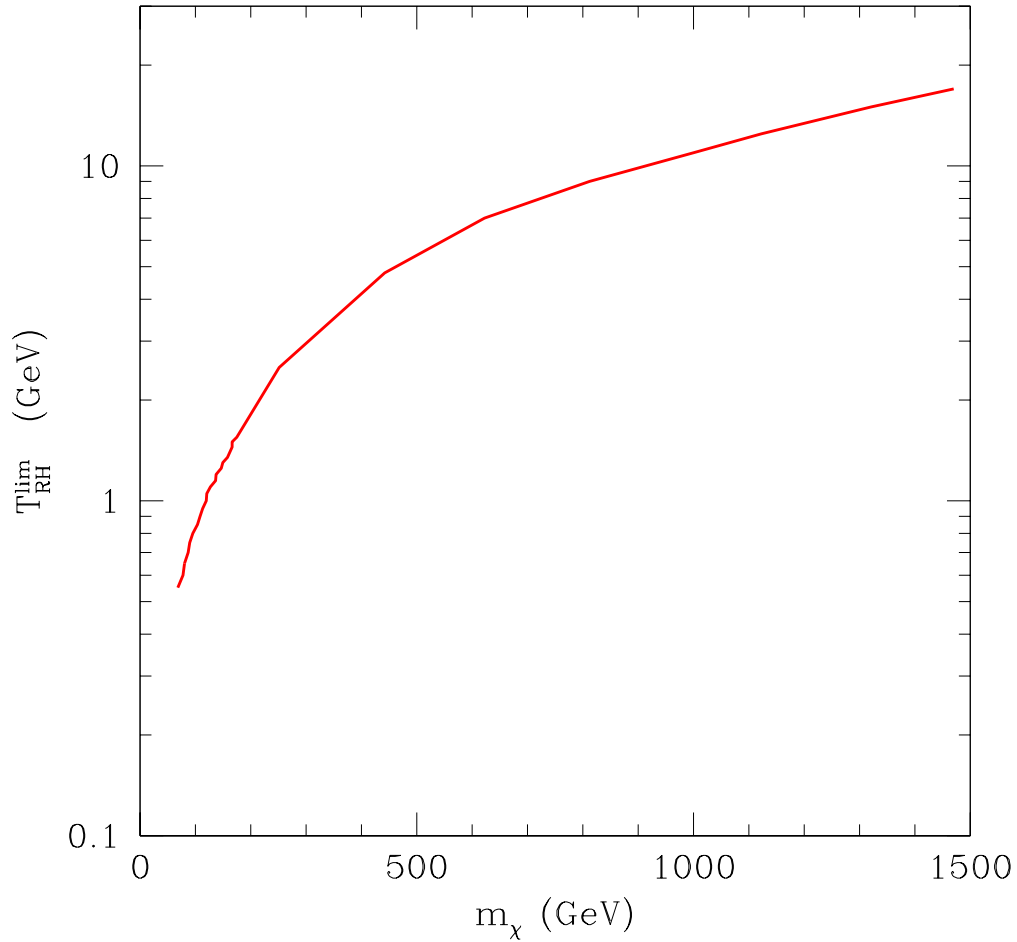} 
\vspace{-50pt}
\caption{Lower
  limit on the reheating temperature $T_{RH}$ as a function of the
  neutralino mass $m_{\chi}$, obtained by requiring that the
  neutralino is the dominant component of dark matter in the Universe
  ({\em i.e.}: $0.05 \leq \Omega_\chi h^2 \leq 0.3$). The result refers
  to universal SUGRA with the parameters varied as in
  Eq. (\ref{eq:ranges}). The values of $m_t^{\rm pole}$, $m_b$ and
  $\alpha_s$ are varied inside their $2\sigma$ allowed intervals.
\label{fig:trh_lim}}
\end{figure}

\clearpage

\end{document}